# Predicting Virtual Learning Environment adoption: A case study

Purpose – To qualify the significance of Rogers' Diffusion of Innovations (DOI) theory with regard to Virtual Learning Environments.

To apply an existing Diffusion of Innovations instrument on a case organisation, the Royal University of Bhutan (RUB), in order to compare its results with previous findings. Descriptive statistics and logistic regression analysis were deployed to analyze adopter group memberships and predictor significance in Virtual Learning Environment adoption and use.

Findings – The Diffusion of Innovations theory is not stable across organizations when it comes to predicting different user categories or the distribution of users. However, it was possible to achieve reliable results for virtual learning environments within a particular organization.

Research limitations/implications – The study questions scholarly attempts to establish models of this type across organizations.

Practical implications – Professionals should be aware that cross-organizational generalizations from Diffusion Of Innovation findings within the domain of virtual learning environments may be very unreliable.

Originality/value – The study challenges the massively cited Diffusion of Innovation literature. It provides data from Bhutan, which is underrepresented in empirical investigations.

## Introduction

The diffusion and adoption of Information and Communication Technologies (ICT) have created an opportunity for universities to complement traditional face-to-face classroom teaching. Moreover, adoption and effective utilisation of ICT in education have become an acknowledged issue of strategic importance in educational institutions around the world (Jebeile & Reeve, 2003). The diffusion of innovations is happening across the globe, and has resulted in adoption or rejection, depending upon the users' perception of the innovations.

The adoption of innovations in higher education can be explained through Rogers' theory of the Diffusion of Innovations (sometimes DOI) (Rogers, 2003). Rogers's theory is widely used as a framework for technology adoption and is composed by a number of factors that influence the motivation of users to facilitate the rate of adoption (Sahin, 2006). For instance, VLEs (Virtual Learning Environments), which are the focus of this paper is a technology which practitioners need to advertise, internally promote and disseminate, and Rogers' DOI theory can be very helpful in this regard.  Rogers (2003) states that the rate of adoption by various adopters (teaching faculty) depends on the factors or characteristics of a given innovation, which in our case is VLEs. Jebeile & Reeve (2008) outlined that after the evaluation of those factors/characteristics, it enables education administrators to plan and design educational technology and infrastructures. This adds strategic importance to the evaluation by practitioners of various types of faculty, to determine their readiness, adoption powers and DOI factors in order to provide institutional management with knowledge and adequate monitoring instruments, supporting improved planning. This can be highly useful for targeting training, addressing characteristics of ICT tools, etc.

Previous research (e.g. Al-Ali, 2007; Keesee & Shepard, 2011; Kilmon & Fagan, 2007; Zayim et al., 2006; Naveh, Tubin & Pliskin, 2006) has studied VLE adoption from a DOI perspective, but still as isolated case studies. Graham, Woodfield & Harrison (2013) use DOI for VLEs, but concentrate not on the predictors, but on the process. By providing comparative results of Keesee & Shepard's instrument for educational technology, the present study strengthens the area of knowledge of Diffusion and Innovation in this specific context. However, it is not a study that merely verifies earlier literature; rather, it demonstrates that for VLE, DOI theory may still be precarious. Furthermore, none of the aforementioned case studies were carried out in a developing country. We contribute to this research area by analysing an interesting case, Royal University of Bhutan. It is the only major university of the country, and its activities are widely dispersed.

In the following sections, we report the academic staff (henceforth, staff) characteristics and the current adoption status, which allows for a logistic regression analysis of these variables. This can be used for the prediction of adoption, and our results indicate that there is more opportunity for prediction than previous literature has found. There are also large variations within the university and between universities in the domain of VLEs, which has significant consequences for other tertiary educational institutes that rely on the DOI literature.

## Current Situation of Royal University of Bhutan

In this section, we describe the research context. RUB is a federated public type University with 10 member colleges distributed across the country. At the time of writing, it has 483 staff and approximately 10,000 students (RUB Statistics, 2013). The RUB ICT strategic plan (Reid & Cano, 2005) has outlined clearly the needs and strategy to set up VLE as one of the components at RUB which can fulfil the expectations at the university of a high level of student involvement and self-learning. This has formed the platform for the integration of technology for teaching and learning. A Wide Area Network (WAN) that connects all the member colleges and server rooms provides the infrastructure for the VLE. It has power backup in order to cope with electricity shortages and malfunctions. The Moodle open source software has been adopted by RUB as its VLE, and has been installed and configured in each college location. RUB formally launched e-learning in Bhutan in May 2011, although Samtse College of Education had been using it since 2004. RUB has adopted a hybrid or blended learning method (Rennie & Mason, 2007), and the VLE facilitates both face-to-face and pure online learning. It provides opportunities for the students to spend less time in a class and engage more in self-directed learning online. Although all the member colleges are connected, RUB is unable to fully support the information exchange on the current bandwidth, which is limited and still lacks consistency to allow full reliance on university-wide solutions (Rennie & Mason, 2007). This is a common state of affairs in developing countries today. Thus, the campus-based setup of VLEs in each individual location was accepted to reduce the pressure on university-wide bandwidth.

Teaching staff at RUB have a great degree of influence on the students, since students relocate to the college and teachers teach their chosen subjects (Rennie & Mason, 2007). For this reason, they are considered to be the ultimate stakeholder group for the future sustenance of the VLE, by encouraging and motivating students to adopt and utilize it until student adoption is total. Even after three years of formal introduction of the VLE facilities, it has been observed that the rate of VLE adoption is very low among the colleges/institutions within the university. Hence, it was found to be necessary to evaluate the level of VLE utilization by staff around the colleges. Despite the training provided to around sixty percent of staff, the number of modules integrated into the VLE is far lower than modules taught only face-to-face. It has become very important to investigate predictors that influence the staff attitude towards the utilization and adoption of e-learning in order to use these to determine the sustenance of VLE.

## Related Work

### Diffusion of Innovation

The probability of new ideas being adopted or abandoned by members of a given culture in the social system is explained by Rogers' (2003) Diffusion of Innovations (DOI) theory. An innovation is defined as an idea, practice, behaviour or object that is perceived by the individual to be "new" (Rogers, 2003). Diffusion is *"the process to communicate an innovation through certain communication channels over time among the members of a social system"* (ibid, p. 5). In the context of the present research, the innovation in question is VLEs, which at this stage is a commonly known acronym at all the Colleges. The diffusion of an innovation is a continuous process that can be examined, facilitated, decided and promoted (Keesee & Shepard, 2011). The rate of adoption of innovations varies, depending upon the innovation types, opinion leaders and types of adopters. Therefore, the DOI theory provides the framework to analyse patterns of staff technology adoption in higher education (Zayim et al., 2006). It states that the technology is not adopted by individuals in the social system at the same time, but this depends on the attitude of the population that has been divided into five categories. The details of these five categories are illustrated in Figure 1.

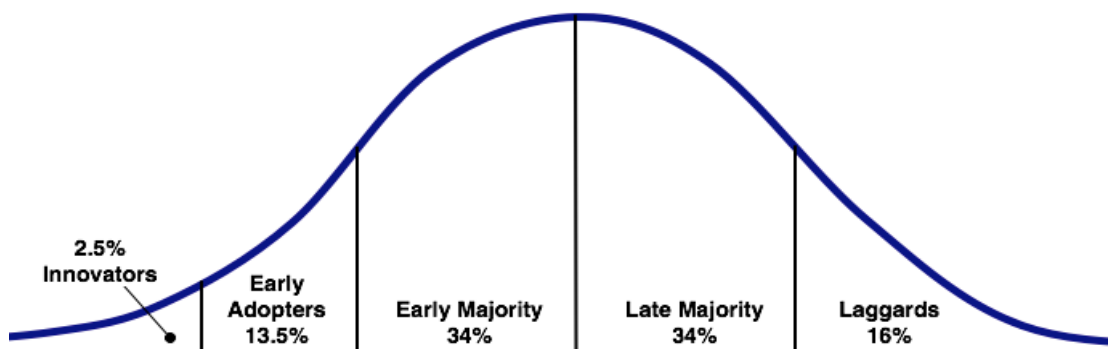

*Figure 1.* Various Adopters (Rogers, 2003, p. 655)

The following segments detail the various adopters as per Rogers (2003):

*Innovators* are the venturesome who are interested in the technical aspects, and are risk takers.

*Early Adopters* are respected and considered as change agents with the greatest degree of opinion about the new ideas. They examine the innovation as regards its benefits and are willing to try it out, provide help and advice to other adopters.

The *Early Majority* is deliberate and more concerned with professionalism. They are willing to adopt the innovation once the majority in society has adopted it.

The *Late Majority* is sceptical and believes less in new ideas and always makes sure that there are people ready to solve their problems before adoption.

*Laggards* are most likely to stick to the "old and traditional" ways. They are very critical towards adopting new ideas, and innovation is accepted only if it becomes tradition.

As Rogers is sometimes misread to contain very simplistic notions of human and social reality, some reservations are warranted here. We do not see these traits as general personality features, but as occurring in the context of the innovation in question, in this case instructional technology and pedagogy. A person can be quick to adopt new customs generally, but a laggard at work; religiously conservative, but an innovator of pedagogy, etc. Drawing the boundary between the adopter categories is an arbitrary act, which is useful to create a common frame of reference for the discussion of diffusion, but the bell-shaped distribution is reported to be a stable empirical finding (Rogers, 2003). It is also important to bear in mind that this is a model which simplifies much more complex patterns in which agency and stakeholdership are distributed in ways that do not fit into the DOI theory. While our basis for this paper is the DOI theory, we are not implying that this can serve as the master frame within all technological-pedagogical innovation , nor do we discuss the merits and limitations of the explanatory power of Rogers' DOI theory. Our specific aim is to refine a model (based on Keesee & Shepard's work) with predictive power for VLE adoption and diffusion and to investigate how this fits in with previous applications. Sometimes predictive models are very useful.

Rogers (2003) identified five attributes that influence attitudes or decisions of an individual during the innovation adoption process. He also claims that those attributes are derived from maximum generality and succinctness and based on past writings and research; they are conceptually distinct, but somewhat interrelated empirically. They influence the likelihood that teachers use the VLE for their daily teaching/learning practice (Askar et al., 2006). The five attributes are relative advantage, compatibility, complexity, trialability and observability:

Relative advantage: The individual considers the current practice and to what degree the innovation would provide advantage. This entails costs and benefits in terms of quality, efficiency, reliability and economic viability – will the adoption of the innovation lead to exceeding the status quo?

Compatibility: Degree of accordance with the existing values, past experiences and requirements of potential users. The innovation should be compatible with the organisational or professional norms or compatible to user needs, social values, standard and ways of working.

Complexity: Degree of difficulty in understanding or using the innovation. The more effort and considerable time it requires, the more unlikely it is that users may adopt it.

Trialability: The perceived possibility to experiment and test the innovation on a limited basis to allow users to understand the benefits of it. If new ideas can be experimented with, this provides ways to the innovators of gaining more understanding of its functionalities on their own terms.

Observability: Degree of visibility to others of results of an innovation. This allows users to observe results and disseminate them to others. The more difficult it is to observe and describe an innovation, the higher is the risk of hindering its adoption. The results can be used to show the effectiveness of using the VLE.

Rogers (2003, p. 298;ibid, p. 316) claims that generally, relative advantage and compatibility are the most important predictors. The perceived attributes or characteristics of innovation predict the rate of adoption among the five group memberships (Rogers, 2003), and the adoption rate is measured as the number of individuals who adopt a new idea in a specified period. Some work is being done on the diffusion of VLEs at the organisational level, but not much at the individual staff member level. With little empirically based research on VLE adoption, we have to assume that with regard to adoption, the staff distribution largely corresponds to the general distribution that Rogers described (later, we will show that this cannot be safely assumed). Zayim et al. (2006) claim that predictors for early VLE adoption (an important user category, since it is key to attaining critical mass) include "non-professorhood" and a high level of self-efficacy. Rogers (1975) has also made a smaller study on instructional innovation in tertiary education, showing (again) that relative advantage is important, as well as observability and trialability, but not compatibility or complexity. He did not study this at adopter category level, which is one of our major foci in this paper.

Table 1
*Predictors for the adopter category of VLEs (Keesee & Shepard, 2011)*

| Category | Predictor |
| --- | --- |
| Innovators | Compatibility and Complexity |
| Early Adopter | Relative Advantage, Complexity and Observability |
| Early Majority | Complexity |
| Late Majority | Compatibility, Complexity, Trialability and Observability |
| Laggards | Relative Advantage, Compatibility, Complexity and Organisational Support |

When the VLE or new features are rolled out, it is important to know which users are innovators and early adopters, since these will diffuse the innovation to the remaining social system. Keesee and Shepard (2011) have developed a predictive model (Table 1) specifically for VLEs. Later, we will examine closely how stable this model is by focusing on our case organization.

## Method

The research method was based on quantitative study collecting demographic information and user perceptions, with minor qualitative supplements (not reported directly in this paper).The intended participants were teaching staff of the Royal University of Bhutan.

Statistical analysis was carried out by using the SPSS package. Descriptive statistics were used to provide patterns of adoption, and logistic regression was deployed to predict the types of staff under various adopter categories.

### Participants

The targeted participants were the full-time as well as visiting staff at various colleges under the Royal University of Bhutan. The study also targeted colleges that offer courses through local language instruction as they also offer programs through the VLE. As a result, it was expected that we would receive acceptable response rates from all major categories of respondents.

Around sixty percent of the total staff have been trained in producing courses with good usability, and in the administration and management of the VLE. However, colleges such as Samtse College of Education (SCE) had prior knowledge on its use as the majority of its faculty members have been providing distance education programs to in-service candidates. Participants in this survey were focussed on teaching staff since they are the ultimate users creating and enabling a platform for their own subject by adding materials and learning activities to be offered for the continuous use of their students. Thus, adequate faculty participation is clearly a critical success factor of the VLE. This is expected to match the requirements outlined in the criteria specified in this study. The specific criteria requirements for participants are illustrated below:

- Full-time teaching staff VLE administrators (they are either teaching staff or members of staff responsible for the VLE)
- Adjunct staff who have the same access rights as regular staff, although they are less exposed to training and less familiar with the ICT facilities.

### Instruments

The quantitative instruments providing the demographic information were developed from the Diffusion of Innovation theory (Rogers, 2003) and in particular Keesee and Shepard (2011). Samples were thoroughly discussed with the Directors, Deans of Academic Affairs and non-IT personal to match the level of understanding of RUB staff as VLE users. All questionnaires were in English (all staff have competency in English to read and write) and divided into three parts, Part 1, Part 2 and Part 3. Part 1 focused on the demographics of respondents, training and experience in VLE, frequency of using VLE, number of modules uploaded and VLE features used for their uploaded online module. Part 2 was based on the 4-point Likert-type scale with a scale range of 1 to 4 to rate their perceptions between two extremities: 1 (*Strongly Agree*) and 4 (*Strongly Disagree*). This is a small improvement of the original Keesee & Shepard instrument, developed in order to force choice (see also Clason & Dormody, 1994). Part 3 contained open-ended questions to enable the respondents to provide their suggestions and comments for future improvements.

Cronbach's alpha test was applied to assess the internal reliability of instruments. The individual predictors' internal reliability are as follows:

Relative Advantage: 0.770
Compatibility: 0.975
Trialability: 0.890
Observability: 0.792
Complexity: 0.682

The reliability analysis of the overall instruments showed a Cronbach's Alpha of 0.934. A validity test was not conducted as all dependent and independent samples were adopted from Keesee and Shepard (2011) who already deployed it.

**Data Collection**

Data were collected from January 2013 to 30 August 2013, both online and by hardcopy questionnaires. The online questionnaire was developed using Google apps and was distributed by email to the following Colleges:

1. Sherubtse College (SC)
2. Samtse College of Education (SCE)
3. Paro College of Education (PCE)
4. Gaeddu College of Business Studies (GCBS)
5. College of Science and Technology (CST)
6. College of Natural Resources (CNR)

At the same time, printed copies were distributed to Jigme Namgyel Polytechnic (JNP) – 30 copies, Institute of Language and Culture Studies (ILCS) – 25 copies, Royal Institute of Health Sciences (RIHS) – 20 copies, National Institute of Traditional Medicine (NITM) – 10 copies, College of Science and Technology (CST) – 30 copies, 20 and 30 copies to Paro College of Education (PCE) and Sherubtse College Education (SCE) respectively. Out of these, 58 stakeholders (GCBS – 33, CNR – 19 and Sherubtse College – 6) responded online and, 143 ( PCE – 25, NITM – 9,RIHS – 17, Sherubtse College – 20 and JNP – 14) were received by post.

## Results and Analysis

In total, 201 staff members participated from all Colleges of RUB, resulting in a response rate of 41.61% out of 483 staff, including expatriates. Thus, the sample collected was considered sufficiently representative for this research study. Out of the 201 respondents, 46 (22.9 %) were female and 155 (77.1%) were male. The age group '41 years and above' is the single largest group with 62 (30.8%) respondents. CST and GCBS had the highest response rate (33 or 16.4%), followed by SC and PCE 25 (both on 12.4%) . The response rate at SCE was 19 (9.5%), at ILCS 18 or 9.0%, at RIHS 17 or 8.5%, at JNP 14 or 7.0 % and at NITM 6 or 3.0%.

Table 2
*Demographic distribution (N = 201)*

| Items | Group | Frequency | Percentage (%) |
|---|---|---|---|
| Gender | Female | 46 | 22.9 |
| | Male | 155 | 77.1 |
| Age | 20 to 25 years | 11 | 5.5 |
| | 26 to 30 years | 43 | 21.4 |
| | 31 to 35 years | 45 | 22.4 |
| | 36 to 40 years | 40 | 19.9 |
| | 41 and above | 62 | 30.8 |
| College | SC | 25 | 12.4 |
| | CST | 33 | 16.4 |
| | JNP | 14 | 7.0 |
| | CNR | 11 | 5.5 |
| | ILCS | 18 | 9.0 |
| | RIHS | 17 | 8.5 |
| | NITM | 6 | 3.0 |
| | PCE | 25 | 12.4 |
| | SCE | 19 | 9.5 |
| | GCBS | 33 | 16.4 |
| Training | No | 69 | 34.3 |
| | Yes | 132 | 65.7 |
| Duration | No training | 69 | 36.3 |
| | 1 to 3 days | 95 | 45.3 |
| | 4 to 6 days | 10 | 5.0 |
| | One week and above | 27 | 13.4 |

It was found that a total of 132 or 65.7 % of participants have been trained in administration and the management of Moodle. Training was conducted at the respective college premises with varying duration from 3 days to more than a week. 27 participants received training for more than a week, and 10 attended 4 to 6-day courses. Table 2 shows the demographic backgrounds of the participants in detail:

**Frequency of usage by staff members**

Table 3 shows the frequency of VLE use in daily teaching and learning.

Table 3
*Frequency of using VLE*

| College | | | How often do you use VLE for your teaching and learning? | | | | | |
|---|---|---|---|---|---|---|---|---|
| | Missing | Daily | Once a week | Once a month | Once a semester | Only once | Never | Total |
| SC | 2 | 3 | 9 | 6 | 2 | 2 | 1 | 25 |
| CST | 2 | 4 | 13 | 7 | 3 | 4 | 0 | 33 |
| JNP | 0 | 1 | 6 | 2 | 0 | 1 | 4 | 14 |
| CNR | 0 | 2 | 2 | 3 | 1 | 1 | 2 | 11 |
| ILCS | 2 | 1 | 8 | 2 | 0 | 3 | 2 | 18 |
| RIHS | 4 | 1 | 5 | 4 | 1 | 1 | 1 | 17 |
| NITM | 1 | 0 | 1 | 4 | 0 | 0 | 0 | 6 |
| PCE | 2 | 2 | 12 | 5 | 0 | 2 | 2 | 25 |
| SCE | 0 | 8 | 8 | 1 | 1 | 1 | 0 | 19 |
| GCBS | 0 | 0 | 3 | 6 | 0 | 9 | 15 | 33 |
| Total | 13 | 22 | 67 | 40 | 8 | 24 | 27 | 201 |

The above indicates that around 22 participants have used the VLE only once, and around 27 (12.9%) have not used it at all in teaching and learning. There are missing values (around 7.4 % of respondents) and some apply to system administrators in the colleges, since these do not fall under the teaching category. However, the interpretation of the work of Choeda et al (2014, p. 214), which uses a different data set, largely corroborates the distribution on the RUB level as a whole.

**Utilisation of VLE functionalities**

Different Moodle functionality is used for the delivery of different teaching contents such as materials, audio visuals, online assignments, grading, forums, online surveys, interactive courses and resources developed using multimedia tools (including more interactive content). Table 4 shows the various existing functionality deployed on courses in the respective colleges. Uploading documents (word, pdf, etc.) has been adopted most widely at CST.

Table 4
*Moodle features deployed*

| College | Material (Word, Pdf, etc) | Audio/ Videos | Assignments | Online Grading | Forums or Chats | Surveys | Multi-media tools (more interactive) |
|---|---|---|---|---|---|---|---|
| SC | 21 | 2 | 17 | 1 | 8 | 2 | 1 |
| CST | 31 | 1 | 24 | 13 | 4 | 4 | 0 |
| GCBS | 12 | 4 | 10 | 1 | 2 | 2 | 1 |
| PCE | 22 | 4 | 12 | 6 | 10 | 1 | 1 |
| SCE | 17 | 11 | 19 | 13 | 11 | 3 | 3 |
| ILCS | 12 | 3 | 13 | 4 | 1 | 0 | 0 |
| RIHS | 11 | 1 | 8 | 3 | 4 | 0 | 0 |
| NITM | 4 | 0 | 4 | 1 | 2 | 0 | 0 |
| JNP | 14 | 1 | 13 | 4 | 1 | 0 | 0 |
| CNR | 8 | 3 | 6 | 3 | 3 | 0 | 0 |

Online assignment is becoming common in the colleges. However, RUB staff is less prone to use interactive contents involving multimedia tools, even though these are used by a few colleges (SC PCE, SCE and GCBS). Other features such as Forums, online grading and audio/video are also used by most of the Colleges. Figure 2 illustrates the overall usage of Moodle activity for the delivery. The uploading of simple teaching/learning material (word, PDF, etc.) features is the most commonly implemented at all the colleges (29.93%), followed by online assignments with 36.10 %. The lowest usage is interactive contents using 'multimedia tools' (1.43%), followed by online Surveys (2.85%). Other functionalities have somewhat similar rates; Audio/Video Teaching (7.13 %), Online Grading (11.64%) and Forums/Chats (10.93%). These data show that while we will later see the VLE as a whole in the later logistic regression analysis, it is a simplification; a VLE is actually adopted in parts. Overall, these data do not indicate successful complete adoption of the VLE, but only partial adoption, with some colleges having significantly lower levels of use.

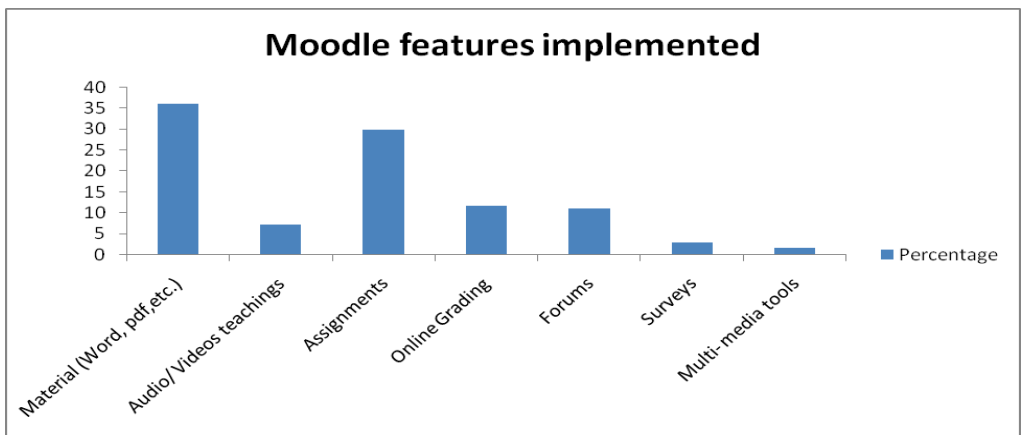

*Figure 2*. Overall percentage of features used

**Predictors for Adoption**

Table 5 (next page) lists some of the instruments derived from Keesee & Shepard (2011) to assist researchers in identifying actual learning activities deployed by academic staff. These instruments were also used to calculate the adopters' status at ten RUB colleges. The right column in the Table 5 was used to categorize the adopter's group by assignment of dichotomous variables (0, if it does not belong, and 1, if it falls under that particular category). The respondent was asked 3-4 questions with dichotomous answers. The respondents were categorized as the adopter group with the highest resemblance to the

Table 5
*Instruments for verifying the categories of staff*

| Instruments | Adopter Type |
|---|---|
| I try new available features of VLE on my own. | Innovators |
| I try new VLE features with the aim of improving teaching and learning. | |
| I share my experience of VLE with my colleagues. | Early Adopters |
| My colleagues often ask me for help to solve VLE problems. | |
| I am using VLE after evaluating its value. | Early Majority |
| I make sure that the VLE for my module is free of problems. | |
| I make sure that I have the necessary technical support to use VLE. | |
| I am not convinced about the value of VLE in my teaching. | Late Majority |
| I started using VLE when the majority of the staff started using it. | |
| I use VLE only when it is necessary. | |
| I do not use VLE for my teaching. | Laggards |
| I am not interested in using VLE for my teaching. | |
| I think VLE will make my teaching worse. | |
| I do not use VLE as my teaching works well without. | |

stereotype (i.e. agreeing to all statements for that category). Respondents had to rate their attitudes based on the statements related to the predictors. These were later subjected to analysis, in which the attitudes were matched with Rogers' predictor categories. The concept was derived from Keesee & Shepard (2011). The overall distribution of staff categories across the colleges is given below:

Table 6
*Distribution of adopters*

| College | Innovators | | Early Adopters | | Early Majority | | Late Majority | | Laggards | |
|---|---|---|---|---|---|---|---|---|---|---|
| | Total | % | Total | % | Total | % | Total | % | Total | % |
| SC | 3 | 12 | 6 | 16 | 8 | 16 | 4 | 16 | 4 | 16 |
| CST | 5 | 15.15 | 8 | 24.24 | 12 | 36.36 | 7 | 21.21 | 1 | 3.03 |
| JNP | 3 | 21.43 | 1 | 7.14 | 3 | 21.43 | 2 | 14.29 | 5 | 35.71 |
| CNR | 4 | 36.36 | 3 | 27.27 | 0 | 0 | 3 | 27.27 | 1 | 9.09 |
| ILCS | 2 | 10.53 | 7 | 36.84 | 4 | 21.05 | 4 | 21.05 | 2 | 10.53 |
| RIHS | 0 | 0 | 1 | 5.88 | 7 | 41.18 | 5 | 29.41 | 4 | 23.53 |
| NITM | 1 | 16.67 | | 0 | 4 | 66.67 | | 0 | 1 | 16.67 |
| PCE | 2 | 8 | 7 | 28 | 9 | 36 | 4 | 16 | 3 | 12 |
| SCE | 2 | 10.53 | 10 | 52.63 | 6 | 31.58 | 1 | 5.26 | | 0 |
| GCBS | 1 | 3.03 | 1 | 3.03 | 11 | 33.33 | 5 | 15.15 | 15 | 45.45 |
| Total | 23 | 11.39 | 44 | 21.78 | 64 | 31.68 | 35 | 17.33 | 36 | 17.82 |

The Early Majority consisted of 64 staff followed by Early Adopters (44). Late Majority and Laggards had almost the same numbers, with 35 and 36 staff respectively. The Innovators category comprised the lowest, with only 23 staff.

The mean and standard deviations of the attributes or predictors are:

Table 7
*Mean and standard deviations (Stdev) of predictors*

| Predictors | Mean | Stdev |
|---|---|---|
| Relative Advantage | 3.11 | 0.89 |
| Complexity | 2.90 | 0.84 |
| Compatibility | 2.79 | 0.89 |
| Trialability | 2.63 | 0.88 |
| Observability | 2.80 | 0.83 |

Table 7 shows the predictors (independent variables) with the mean and standard deviation calculated. These were considered for the logistic regression to predict the probability of staff categories..The logistic regression analysis was applied to calculate the odds and odds ratio (Exp (B)). The significant predictors (significant value considered was less than 0.05) can predict the likelihood of category membership as provided by Rogers. The predictors are Relative Advantage, Complexity, Compatibility, Trialability and Observability.

Table 8
*Results of the significant predictors for RUB Staff Categories*

| RUB Staff Categories | Relative Advantage | Complexity | Compatibility | Trialability | Observability |
|---|---|---|---|---|---|
| Innovators | Exp(B): 1.624 | Exp(B):0.340 Significant | Exp(B):0.531 | Exp(B): 2.711 Significant | Exp(B): 9.105 Significant |
| Early Adopters | Exp(B):1.680 | Exp(B): 2.467 Significant | Exp(B): 1.174 | Exp(B):0.947 | Exp(B): 0.739 |
| Early Majority | Exp(B):0.698 | Exp(B): 1.537 | Exp(B): 2.224 Significant | Exp(B): 0.781 | Exp(B): 1.041 |
| Late Majority | Exp(B): 0.970 | Exp(B): 0.294 Significant | Exp(B):1.123 | Exp(B): 1.168 | Exp(B): 1.091 |
| Laggards | Exp(B): 0.945 | Exp(B): 1.298 | Exp(B): 0.320 Significant | Exp(B):0.977 | Exp(B): 0.228 Significant |

Table 8 reflects the odds ratio (Exp (B)) that determines the likelihood to accurately predict an adopter category. If an odds ratio is more than 1, this signifies that perceived predictors amongst staff are more likely to belong to that category. However, if it is less than 1, this signifies that perceived predictors amongst staff are less likely to belong to a given category. Space restricts a full explication of how to derive probabilities and other characteristics; but an example of how to translate these results regards the odds for innovators, using the natural logarithm of the standard logistic regression model (see Grimm & Yarnold, 1995). Table 8 shows that the participant perceptions Relative Advantage, Trialability and Observability will result in the odds 1.624, 2.711 and 9.105 to 1 that a VLE user belongs to the innovators category. However, if Complexity (0.340) and Compatibility (0.531) are indicated, the odds of being an innovator are low – and so on for the rest of the items in the table. However, only the significant predictors can be used for our final purposes.

As indicated in Table 8 shows that the significant predictors are Compatibility for Early Majority and Laggards, Complexity for Innovators, Early Adopters, Early Majority and Late Majority, Trialability for Innovators, and Observability for Innovators and Laggards.

**Discussion**

In relation to previous research by Keesee & Shepard these findings can be summed up as follows (Table 9). Table 9 indicates that the predictors are quite different in local contexts. General models for predicting adoption should be used with caution. However, some commonalities were also found. An interesting feature of Keesee & Shepard is that all groups find complexity important. Hence, this is not useful for predicting adopter category, as "complexity" attention of a user cannot be used as a differentiating attribute. In our model, however, the Early Majority and Laggards can be ruled out. There are significant variations between the graph illustrated by the distribution of RUB Adopters and Rogers' bell curve (Figure 2). We see here a deviation in the innovator part of the curves. However, while this result may be important, we are careful when drawing conclusions from it. The distance between the curves is large, but the RUB curve is based on very few innovator-respondents, so there is a risk of a respondent bias.

Table 9
*Our verified predictors as compared with previous research*

| Membership group | Predictor found in both studies | Keesee & Shepard only | Our Present study only |
|---|---|---|---|
| Innovators | Complexity | Compatibility | Triability and Observability |
| Early Adopters | Complexity | Relative Advantage and Observability | No category |
| Early Majority | No category | Complexity | Compatibility |
| Late Majority | Complexity, | Compatibility, Trialability, and Observability | No category |
| Laggards | Compatibility | Relative Advantage, and Complexity | Observability |

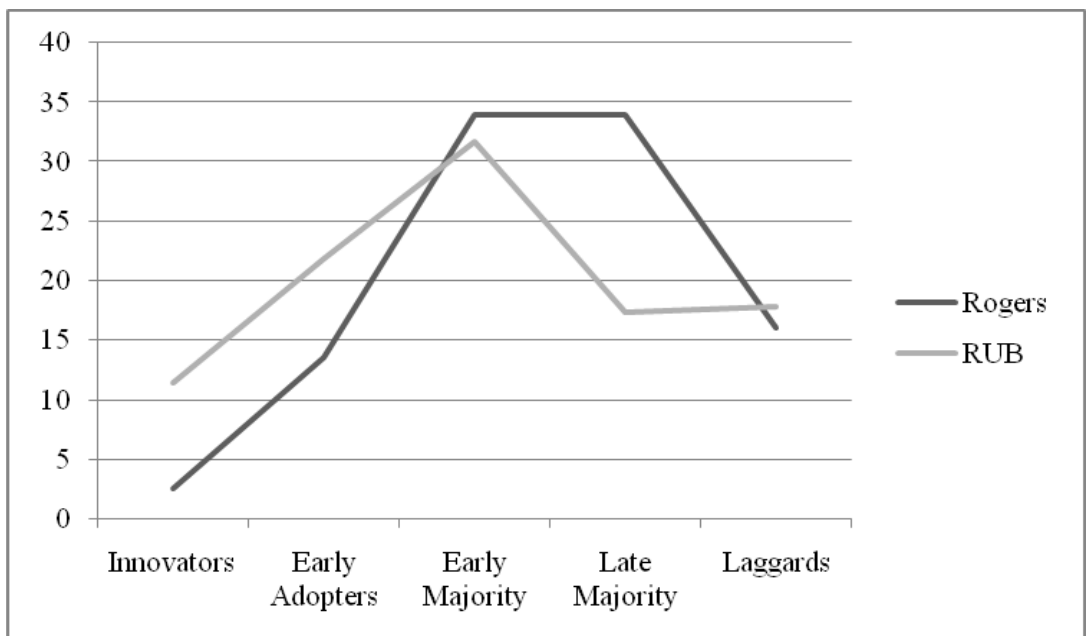

*Figure 2.* Variations in adopter categories

Table 10
*Variation of adopter categories*

|  | Innovators | Early Adopters | Early Majority | Late Majority | Laggards |
|---|---|---|---|---|---|
| Rogers (standard Bell curve) | 2.5 | 13.5 | 34 | 34 | 16 |
| RUB | 11.4 | 21.8 | 31.7 | 17.3 | 17.8 |

Table 10 shows that the distributions of adopters at RUB were compared with Rogers, signifying a clear distinction in variations in distribution patterns. It shows that the distribution is more favourable at RUB. The percentage of Innovator and Early adopters is higher than in Rogers, that of Early Major and Laggards is almost equivalent to Rogers, that of Late adopters is lower than Rogers as compared to the predicted power of adoption from Rogers' population distribution. What is more interesting is that a majority of the population (staff around 65%) belonged to the categories of Innovators, Early Adopters and Early Majority at RUB, as compared to 50% in the case of Rogers. This would normally indicate (ceteri paribus) that the organisation easily adopts innovations that they are exposed to, yet it is not the case here, despite training and management support. It is the policy of RUB for each College to upload at least 10 to 20 % modules of the programs (Author, 2011) although no specific encouragement has been given in terms of using interactive modules. (Choeda et al, 2014; Author, ibid) state that most of the teachers and students at RUB perceived VLE as useful as it saves a significant amount of their time and resources and was used to share benefits with other users. It can be inferred from Table 2 that many of the staff who used document uploading have not implemented the more interactive/"advanced" features. The VLE does not appear more complex than comparable Moodle installations (although this study has not formally investigated this). Does lack of training account for the reluctance to adopt VLE? Rogers (2003) explains that the adoption rate depends on the individual's perception and the extent of the 'promotion efforts', and training will naturally change perceptions. Around 40% of the staff have not been trained and, moreover, new staff members are recruited at the beginning of every year. It takes time for them to get accustomed with the educational technology although they have been informed by their college on the VLE. It is difficult to find studies that benchmark VLE training across institutions, but we think that other universities have manage richer and higher adoption rates with less training. Indeed, other studies of VLE use at RUB show that uptake is slow despite training (Kinley, 2010). Rogers (1983, p. 233) gives three other factors of importance:

1. Type of innovation-decision, where 'authoritative decisions' are the fastest. RUB's adoption of the VLE is championed by the top management and is a kind of authoritative decision, yet it has not assisted the process.
2. Communication channels. We have no data in this regard; this *remains a possible explanation to the slow adoption process*.
3. Nature of the Social system. This is *also a possible explanation*. RUB's existing norms, degree of interconnectedness, etc., may be impeding the processes.

This shows that adopter frequency and perceptions do not show the full picture of adoption, as Rogers concedes (but sometimes the other factors above are overlooked in the model; Keesee & Shepard is an example of this). It also shows that adoption in the VLE case is not an issue of adopting the *whole* of an innovation (which is also briefly mentioned in Rogers (1975)). If we look at the VLE as a tool for the distribution of PDFs to the students for their course, Rogers' model has more (but still not good) explanatory power.

There are alternatives to Rogers' explanation, such as Moore and Benbasat's refined instrument (1991). Another intellectual option is to abandon the attempt to establish general models. Some studies go for in-depth studies of VLE adoption (Nyvang, 2008), and typically find additional case-specific variables, rather than the 'universal' predictors. In-depth studies also reveal whether the VLE is a non-changing unit of analysis, or if users gradually start to perceive it not as one VLE-object, but as several, or in a qualitatively new way. It is not the objective of this paper to show the merits of the alternatives exemplified above, but further research may fruitfully compare them in the area of VLEs.

## Conclusion

The findings of the study reveal that perceptions of predictors by academic staff determine the likelihood of belonging to a certain group of VLE adopter, e.g. Early or Late Majority. RUB has implemented VLE to enhance the current traditional types of learning. The study reflected that the utilization of VLE is not particularly satisfactory in terms of deploying the interactive contents. The regression analysis shows that RUB diverges from previous research in terms of the prediction as to which adoption type staff belong to. This means that to generalizing findings across institutions and innovations within the area in question will be ill-founded. Rogers was also painfully aware of the limitations of his own approach (see Rogers, 1983, p. 130ff). Our work provides empirical ground for the many conceptual critiques (Schön, 1973; Lundblad, 2003; see Denning (2010) for a good introduction) of Rogers. Nevertheless, our study also shows that it is possible to build a local theory of adoption of VLEs that can be useful for RUB itself. Accordingly, it may be fruitful for other institutions to apply the instrument from this article, and to derive an equivalent model, based on their own data. Another route is to make the instrument more comprehensive in order to achieve cross-institutional generalizability. Keesee & Shepard do not take all factors of rate of innovation into account. It could be interesting to add these (innovation decision, communication channels, and the nature of the social system) to the instrument, or to investigate which complementary research tools would cover these factors satisfactorily.

Furthermore, universities should be aware of the fact that the adoption distribution is far from uniform within the organisation, and that it may not predict the adoption very well at college-level. Some colleges have large bases of early adopters. A diversified strategy for broadening the user base seems important, as the case of RUB shows. In some colleges, the majority of adopter groups are under the category of Late Majority and Laggards, which signifies that the college management or RUB need to offer more assistance to them and add more importance to the significant predictors that can help them forecast the adoption of VLE amongst academic staff, as well as group memberships. This applies in particular to Late Majority and Laggards to make sure that they do not remain undetected.